\documentclass[aps,pra,twocolumn,superscriptaddress]{revtex4-1}
\usepackage[utf8]{inputenc}
\usepackage{graphicx}
\usepackage{amsmath,amssymb,amsfonts}
\usepackage{bbold}
\usepackage{enumitem}
\usepackage{xcolor}
\usepackage{soul,xcolor}
\usepackage{tikz,tkz-euclide}
\def\DJo{$\;$\kern-.4em \hbox{D\kern-.8em\raise.15ex\hbox{--}\kern.35em okovi\'c}}
\def\CC{{\rm\kern.24em \vrule width.04em height1.46ex depth-.07ex
\kern-.30em C}}
\def\P{{\rm I\kern-.25em P}}
\def\NN{{\rm I\kern-.15em N}}
\def\RR{{\rm
         \vrule width.04em height1.57ex depth-.0ex
         \kern-.03em R}}
\def\RR{{\rm I\kern-.23em R}}
\def\id{{\rm 1\kern-.22em l}}
\def\ZZ{{\sf Z\kern-.44em Z}}

\newtheorem{psatz}{Satz}[section]
\newtheorem{pdef}{Definition}[section]
\newtheorem{conjecture}{Vermutung}[section]

\newenvironment{eqblock}[2]{\beq\label{#2}\begin{array}{#1}}{\end{array}
                                \eeq}
\newenvironment{neqblock}[1]{\[\begin{array}{#1}}{\end{array}\]}
\newcommand{\beqb}{\begin{eqblock}}
\newcommand{\eeqb}{\end{eqblock}} 
\newcommand{\nbeqb}{\begin{neqblock}}
\newcommand{\neeqb}{\end{neqblock}}

\newcommand{\eps}{\varepsilon}
\newcommand{\beq}{\begin{equation}}
\newcommand{\beqa}{\begin{eqnarray}}
\newcommand{\eeq}{\end{equation}}
\newcommand{\eeqa}{\end{eqnarray}}
\newcommand{\nbeqa}{\begin{eqnarray*}}
\newcommand{\neeqa}{\end{eqnarray*}}

\newcommand{\ket}[1]{| #1 \rangle}
\newcommand{\ketbra}[1]{\ensuremath{| #1 \rangle \langle #1 |}}

\newcommand{\Matrix}[2]{\left( \begin{array}{#1} #2 \end{array}
  \right)}

\renewcommand{\Re}{{\rm Re\;}}

\usepackage[colorlinks=true,citecolor=blue,linkcolor=blue,urlcolor=blue]{hyperref}

\usepackage{graphicx} 
\begin{document}

\date{July 2024}
\title{The exact convex roof for GHZ-W mixtures for three qubits and beyond}

\author{Andreas Osterloh}
\affiliation{Quantum Research Center, Technology Innovation Institute, Abu Dhabi, P.O. Box 9639, UAE}

\begin{abstract}
I present an exact solution for the convex roof of the square root of the threetangle for rank two density matrices and for all states within the Bloch sphere.
Aside the formerly known two tetrahedra it contains three additional optimal tetrahedra and connecting triangular optimal decompositions. The remaining optimal decompositions are one-dimensional. Optimal decompositions are proved to contain as many states from the zero-polytope as possible, a property that is called zero-state locking; it will be the working horse throughout this work. In addition, an inequality is derived which decides about the optimality of the decompositions under consideration. The footprint of the measure of entanglement consists in a characteristic pattern for the fixed pure states on the Bloch sphere surface which constitute the optimal solution. This solution is subject to transformation properties due to the SL-invariance of the entanglement measure which renders the optimal decomposition found here to all the states within the SL-class of GHZ and W. The method presented here is directly applicable to the symmetric mixture of generalized GHZ and W states for arbitrary number of qubits but the main structure of the 3-dimensional tetrahedra is general for all rank-two mixtures of states.
\end{abstract}

\maketitle

\section{Introduction}

Entanglement is omnipresent in the presence of interaction and it is difficult to measure, because the operator to measure it is non-Hermitean, a case rarely encountered in physical systems. Its expectation value is therefore not calculated as the simple trace with the density matrix of the state in consideration; instead it is complicated by the convex-roof\cite{MONOTONES} which in general is NP-hard to calculate\cite{gurvits2003classical,Gurvits2004classical}. It is, however, an important quantity, indispensable for quantum computing\cite{ladd2010quantum,gill2022quantum,orus2019quantum,ringbauer2022universal}, quantum communication\cite{gisin2007quantum,luo2023recent,cozzolino2019high,de2024parallel}, and quantum sensing in general\cite{narducci2022advances,degen2017quantum,crawford2021quantum,wang2023photonic,aslam2023quantum,schnabel2010quantum,pezze2018quantum,polino2020photonic}; quantum technology works because of these nonlocal quantum correlations. As a non-local quantity, entanglement has to be invariant under the change of the local basis and permutations of the local entities. This leads to the concept of LOCC invariance and the monotone property\cite{MONOTONES} that every good entanglement measure must satisfy. 
In general, this monotone property needs a proof, which however is granted for by extending the local invariance group from $SU(n)$ to $SL(n)$\cite{VerstraeteDM03NormalForms} for $n$ being the number of different axis of the local entity one can do rotations about. The quantities obtained through this extension may have included the elegant property of being essentially susceptible to entanglement at a global level only, discarding everything which is non-global. These are called measures of genuine multipartite SL-entanglement\cite{OS04,DoOs08,neveling2024threetangle}. Simple examples for qubits as local entities are the concurrence\cite{Hill97,Wootters98} and the threetangle\cite{Coffman00}. The name {\em tangle} is here reserved for general measures of genuine multipartite SL-entanglement, {\em n-tangle} if we want to specify that it consists of n constituents.

\begin{figure}[h!]
    \centering    
    \includegraphics[width=\linewidth]{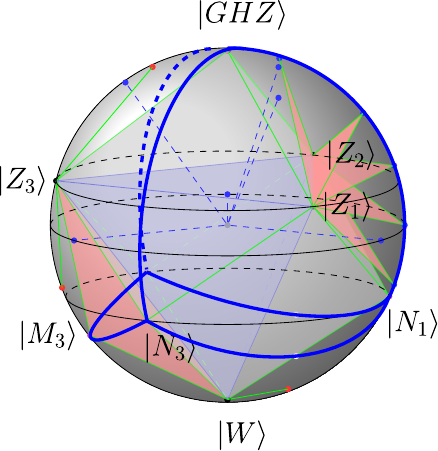}
   \caption{Optimal decompositions for the rank-two mixtures of GHZ and W state are depicted. They consist of the zero polytope (blue shaded), 4 polytopes defined by three of the zero states together with a state $\ket{N}$ one of which is the GHZ state itself (grey shaded); all these polytopes are three dimensional. On top of this kernel structure of five three-dimensional polytopes are three lines along grand circles along $\varphi$ direction
   and three lines which are on circles with a distance $0.0711148$ to the center of the Bloch sphere. Along these lines move pure states which each form a two-dimensional optimal $(2,1)$ decomposition (red triangles). Blue points mark the end position of the respective normal vectors of respective circles on the Bloch sphere surface connected by blue dashed lines. All remaining parts of the Bloch sphere are covered by $(1,1)$ decompositions.}
    \label{fig:OptimDecomp-GHZ-W}
\end{figure}
Included in the monotone property is the convex-roof extension to arbitrary mixed states, which can be NP-hard\cite{gurvits2003classical,Gurvits2004classical}.
For very few examples do we have an exact solution for evaluating this convex-roof, as for the concurrence for two qubits\cite{Wootters98,Uhlmann00}. These are examples for which the pure state measure already scales linear in $\rho$; this is also the case if the corresponding $d$-th root of an n-tangle of degree $2d$ is a bilinear functional as the concurrence. This happens e.g. for the threetangle for certain states\cite{neveling2024threetangle} (see also for telescope states in \cite{Eltschka2009}). It is however always beneficial to consider an SL-invariant measure which scales linearly with the density matrix; this implies that every result one has got for the Bloch-sphere picture easily translates to SL-transformations of the states~\cite{viehmann2012rescaling}. The $SL$-invariance of the n-tangle induces transformation properties of optimal decomposition states\cite{viehmann2011polynomial,viehmann2012rescaling} and lifts the results of this work to all states being SL-equivalent to the mixed state in consideration.
This is the ultimate reason why the choice of pure-state measure falls on the $d$th root of an n-tangle of degree $2d$ as entanglement measure; hence it is $\sqrt{\tau_3}$ (see Appendix \ref{app:threetangle}) for the case analyzed in detail here.
One exclusive property of SL-invariant entanglement measures is that they are given as multi-nomials of complex variables $z_i$ whereas it will depend also on $|z_i|$ whenever one has to do with an SU-invariance of bi-degree $(2n-m,m)$\cite{johansson2014classification} with $0<m\leq n$; for $m=0$ we recover SL-invariance. This simplifies significantly the needed calculations for superpositions of states. If the density matrix, as in this work, has rank two then it is made of two eigenstates. Therefore, a pure state with zero n-tangle of degree $2d$ on the Bloch-sphere is given by the $2d$ solutions of a polynomial of degree $2d$\cite{KENNLINIE}. These {\em zero-states} form a polytope with $2d$ corners in the Bloch sphere picture: the {\em zero-polytope}.
In general the zero-polytope consists of $2d$ independent solutions. It is known\cite{LOSU,EOSU,viehmann2012rescaling} for the mixture of GHZ and W states and the square root of the threetangle that the Bloch-sphere contains two tetrahedra: one is the zero-polytope and the other is the polytope made of three of the states on top of the zero-polytope and the GHZ state. In the general setting where the rank is not bound by two, manifolds would substitute the zero-polytopes in this work\cite{KENNLINIE}. These zero manifolds would have to be convexified themselves, which needs an imbedding in a convexly transferred zero-manifold.
It is tempting to minimize the mixed n-tangle in choosing as many states as possible out of this zero-polytope.
It turns out that one further benefit of considering pure state measures of SL-entanglement that scale linearly with the density matrix is that in this case as many states out of it have to be chosen to yeald the optimum. This behavior corresponds in a locking-in of possible zero-states. A proof of this {\em zero-state locking}\cite{osterloh2016exact,neveling2024threetangle}, a crucial ingredient for the results presented, is discussed in the next section.

Optimal decompositions do not intersect and are here limited to at most $4$ states\cite{caratheodory1911variabilitatsbereich,Uhlmann98} (see Appendix \ref{app:optimal-decompositions}), i.e. to at most three dimensional simplices. 
We use the nomenclature $(n_z,n_e)$ from Ref.~\cite{neveling2024threetangle} for identifying the decompositions: $n_z$ is the number of states from the zero-polytope, and $n_e$ that of entangled pure states in the decomposition. 
In Ref.~\cite{neveling2024threetangle} it has been outlined how optimal decompositions will behave for rank-$2$ density matrices. 
In particular, it has been observed that for each three surface elements of the zero polytope there is precisely one entangled pure state $\ket{N}$ forming a $(3,1)$ polytope. Within these polytopes the entanglement varies linearly due to convexification. Hence, three additional pure states $\ket{N_i}$, $i=1,2,3$, need to be found as a convexification point of $(1,1)$ and $(2,1)$ decompositions. This fills out good part of the Bloch-sphere. The tips of each two states $\ket{N_{i_1}}$ and $\ket{N_{i_2}}$ are connected by a line of pure entangled states. They define further two-dimensional $(2,1)$ decompositions in the Bloch sphere. All remaining decompositions will be of the type $(1,1)$. 
This excludes the possibility of optimal $(0,n_e)$ decompositions for $n_e>1$ here. Optimality of $(n_z,n_e)$ could be excluded numerically through case studies only\cite{neveling2024threetangle}.
In section \ref{sec:decomps} an inequality is given in Eq.~\eqref{eq:2-1vs0-2} which decides about where this behavior changes. It is provably not satisfied for the case studied.

This work is laid out as follows. We will start with
the presentation of the proof of the zero-state locking theorem
underlying this work. Next, we introduce the general framework in section \ref{sec:procedure} which can be subdivided into three parts. It starts with a discussion about the specific symmetries and where they come from, and turns into finding the (in general) three-dimensional zero-polytope. Some remarks on the mentioned generalization is highlighted. In subsection \ref{sec:states-Ni}, a discussion of the additional three dimensional polytopes follows. $(2,1)$ and $(0,2)$ decompositions are discussed, obtaining an inequality that decides about its optimality in the consecutive part \ref{sec:decomps}.
Finally, the intersecting optimal lines between the states $N_i$ are obtained in the subsection \ref{sec:states-Mi}.\\
We apply this procedure to $GHZ$-$W$ mixtures of three qubits and the square root of the threetangle. 
The individual steps and calculations needed are supported by the respective Appendices.
The result are the states $\ket{N_i}$ which form optimal $(3,1)$ decompositions and lines interconnecting each pair of the entangled pure states $\ket{N_i}$. The remaining space of the Bloch-sphere is filled with one-dimensional $(1,1)$ decompositions, depicted in Fig.~\ref{fig:OptimDecomp-GHZ-W}. 
The results are discussed in the conclusions,
where also an outlook is presented.

I want to emphazize that in this work we use doubly occurring symbols (for example $z$ as complex variable and for the cartesian coordinates of the Bloch sphere, d for both the degree of a polynomial and a distance) and multiply occurring notions across the vast literature (e.g. tangle is sometimes used for the linear entropy or shadow which have their meaning in the corresponding literature). I nevertheless use them for ease of description and think this should cause no confusion.

\section{Zero-state locking}\label{sec:lock}
We give a proof of the zero-state locking behavior for optimal decompositions. It states that every zero-state visible from a density matrix $\rho$ is within the optimal decomposition. Here a (zero-)state $Z$ is called visible from a state $\rho$ if the connecting line $\overline{Z\rho}$ does not intersect an existing optimal polytope.
In the very same way the shadow of a state can be defined: interpreting states as pointwise light sources and the existing optimal simplices as intransparent because optimal decompositions cannot intersect, this will lead to shadows as dark spaces on the surface of the Bloch sphere, from where this particular state results invisible.
It leaves those states that can be viewed from a density matrix as the only possible pure states for an optimal decomposition.
Let $\tau_d$ be an SL-invariant measure of entanglement ({\em tangle}) of polynomial degree $2d$; we have $\tau_d[\alpha \psi]=\alpha^{2d}\tau_d[\psi]$. Hence its $d$-th root scales 
quadratic in the wave function coefficients as it does for the concurrence. Thus, we take $\sqrt[d]{\tau_d}$ as a proper measure 
because of its linear scaling behavior in the probabilities~\cite{viehmann2012rescaling}.
The equation $\tau_d[\psi]=0$ has for SL-invariant tangles $\tau_d$ exactly $2d$ solutions. Each single root $z_j$
leads to a scaling behavior $(z-z_j)^{1/d}$. Hence multiple roots scale as $(z-z_j)^{m/d}$ if $m$ is the multiplicity\footnote{I emphasize that $z$ is the stereographic projection of the Bloch sphere. In the $z_j$ the tangle $\tau_d$ is represented by
$\tau_d=\prod_{j=1}^{2d} (z-z_j)$. }.
Let for the locking property be $m<d$. We therefore have a strictly concave behavior around $z_j$.
We next consider a decomposition in which some of the decomposition states are chosen to be elements in the zero-polytope.
Let $\rho$ be the density matrix in consideration and let $\vec{x}_0$ correspond to the vector of the density matrix of only zero-states. Its average tangle is
\beqa
T[\rho]&=:&T[\vec{x}_0]=\sum_{i\in I_0} p_{0;i} T[\vec{z}_i] + \sum_{j\in I_e} p_{e;j} T[\vec{e}_j]\\
&=& \sum_{j\in I_e} p_{e;j} T[\vec{e}_j]\ ,
\eeqa
where $I_0$ is the index set of the zero-polytope, $\vec{z}_i$ 
corresponding to the pure zero-states, $I_e$ the index of pure entangled states, $\vec{e}_j$ corresponding to the pure entangled states. 
\beqa
\rho_0&=&\frac{1}{\sum_{i\in I_0} p_{0;i}}\sum_{i\in I_0} p_{0;i} \vec{z}_i \\
\rho_e &=& \frac{1}{\sum_{j\in I_e} p_{e;j}}\sum_{j\in I_e} p_{e;j} \vec{e}_j 
\eeqa
are the related density matrices.
This tangle can be written in terms of two functions: one containing the concave parts stemming from the zero-polytope and an analytic part for the entangled side, $e(\vec{x}_0)$. 
Therefore, we can write
\beqa
\rho&=&(1-w(\vec{x}_0)) \rho_0 + w(\vec{x}_0) \rho_e \\
T[\vec{x}_0]&=& w(\vec{x}_0)e(\vec{x}_0) =: w_0 e_0\ ,
\eeqa
where $w(\vec{x})$ is the weight of the entangled density matrix and $e(\vec{x})$ its average entanglement. Both the weight function $w$ and the average entanglement $e$ are locally holomorphic functions and have a Taylor expansion.  
Now assume that (keeping the density matrix fixed) one solution of the zero-polytope be shifted by a small amount $\vec{\eps}$ on the Bloch sphere surface, described by two variables\footnote{These two variables could for example be $\varphi$ and $\theta$. The stereographic projection would lead to a stretching of $\theta$ by the value $1+|z_j|^2/4$. We, however, prefer to leave it more general.}. The remaining states stay unaltered without loss of generality.
This consequently leads to a change of both the weight function $w(\vec{x}_0+\vec{\eps})=w_0-\vec{w}_1\cdot \vec{\eps} + \eps_i w_{2;ij}  \eps_j+ \dots$
and the entanglement function $e(\vec{x}_0+ \vec{\eps})=e_0+\vec{e}_1 \cdot \vec{\eps} + \eps_i e_{2;ij}  \eps_j+\dots$ such that the average entanglement of the new decomposition results in 
\beqa
T(\vec{x}_0+\vec{\eps})&=&(1-w(\vec{x}_0+\vec{\eps}))e_z |\vec{\eps}|^{m/d} \\
&& \ + w(\vec{x}_0+\vec{\eps}) e(\vec{x}_0+\vec{\eps}) \nonumber \\
&\approx& w_0 e_0 + (1-w_0)e_z |\vec{\eps}|^{m/d} \\
&& \quad + (w_0 \vec{e}_1-e_0 \vec{w}_1 )\cdot\vec{\eps} \nonumber
\eeqa
with $e_z$ a positive prefactor. For $m<d$ in the limit $\eps\to 0$ one finds $T(\vec{x}_0+\vec{\eps})=w_0 e_0 + (1-w_0)e_z |\vec{\eps}|^{m/d}$ which is larger than $T(\vec{x}_0)$ as long as the condition $m<d$ is satisfied.
This property excludes the possibility of optimal $(0,n_e)$ decompositions as long as some of the zero-states are visible from the density matrix in consideration (optimal decomposition polytopes can be viewed as intransparent for density matrices, since optimal decomposition cannot intersect\cite{caratheodory1911variabilitatsbereich,Uhlmann98}).

For $m=d$ one obtains $(w_0  \vec{e}_1-e_0 \vec{w}_1)\cdot \vec{\eps}+(1-w_0)e_z|\vec{\eps}|$. For this to be positive and the optimal decomposition to stay zero-locked gives the condition $(w_0  \vec{e}_1-e_0 \vec{w}_1)\cdot \vec{\eps}<-(1-w_0)e_z|\vec{\eps}|$. Curiously, the {\em integrable} case - where the covex-roof can be obtained exactly - we have $\sqrt[d]{\tau_d[z]}=(z-z_1)(z-z_2)$, hence both zeros in the system have a multiplicity of $d$. Here, however, the inequality has to become an equality because no zero-state locking is observed and the optimal decomposition can be chosen arbitrarily as in the case of the concurrence (see also Ref.~\cite{Uhlmann00}) where decompositions can be made of equally entangled pure states\cite{Hill97,Wootters98}.
Whenever $m>d$, there is no zero-state locking for this particular state from the zero-polytope. 
For the remaining solutions in the zero-polytope it nevertheless applies.

Because optimal decompositions must be continuous (i.e. their corresponding weight functions must behave continuously \footnote{This is essentially the same as non-intersecting of optimal decompositions. In other words: on every line through the Bloch-sphere the entanglement must behave in a continuous manner)}) or equivalently no two optimal decompositions can intersect each other 
(unless the pure states of the decompositions are joined together to give a larger optimal polytope, this zero-state locking behavior excludes optimal decompositions without  states from the zero-polytope, {\em iff} they result visible
from the density matrix in consideration.

\section{General procedure} \label{sec:procedure}
In~\cite{neveling2024threetangle}, much effort has been put into the distribution of optimal decompositions. These findings hold for even more general SL-invariant n-tangles $\tau_n$.
We describe here the general procedure for an SL-invariant n-tangle of degree $2d$ given a rank two density matrix $\rho$ with eigenstates $\ket{\psi_0}$ and $\ket{\psi_1}$. 
At first the zero-polytope has to be determined by solving\cite{LOSU,KENNLINIE} $\tau_n[\ket{\psi_0}+z \ket{\psi_1}]=P_{2d}[z]=0$ with all states whose n-tangle is zero as zero-states. Here, $P_{2d}$ is a polynomial of degree $2d$ of $z$.
The n-tangle is zero within this convex polytope and non-zero outside. The polytope has two dimensional simplices as faces, which comprises three pure states; for each face one has to find the entangled pure state of the optimal decomposition states and the lines interconnecting them. This structure of points and lines is the signature of the optimal decomposition states of the specific n-tangle to all rank-two density matrices composed of the states $\ket{\psi_0}$ and $\ket{\psi_1}$.\\
We here analyze the Bloch sphere of the rank-two mixed state
\beq
\rho[p]=p\ketbra{GHZ} + (1-p) \ketbra{W}
\eeq
where $\ket{GHZ}=(\ket{111}+\ket{000})/\sqrt{2}$ and $\ket{W}=(\ket{100}+\ket{010}+\ket{001})/\sqrt{3}$ and $p$ being a probability of a convex decomposition.
A three-fold rotational symmetry due to cyclic permutations of the qubits exist for both states. More important is however the symmetry due to the n-tangle, which here is the threetangle.
With $\ket{GHZ}=(\ket{000}+\ket{111})/\sqrt{2}$ and $\ket{W}=(\ket{100}+\ket{010}+\ket{001})/\sqrt{3}$ 
being the poles of the Bloch sphere, the pure states' threetangle is
\beqa
\tau_3(p,\varphi)&=& \tau_3\left[\sqrt{p}\ket{GHZ} + \sqrt{1-p}e^{i\varphi} \ket{W}\right]\\
&=& p^2+\frac{2^{7/2}}{3^{3/2}} \sqrt{p(1-p)^3}\, e^{i 3 \varphi}
\eeqa
Therefore,
it is enough to perform the classification in a wedge of the Bloch sphere given by the azimuthal angle $\varphi\in [-\pi/3 , \pi/3]$ only. The rest of the sphere is obtained by symmetry requirements.
There are three additional zero-states $\ket{Z_j}$, $j=1,2,3$ were obtained as $\ket{Z_j}=\sqrt{p_0}\ket{GHZ}-\sqrt{1-p_0}\, e^{i\frac{2\pi (j-3)}{3}}\ket{W}$ with $p_0=\frac{4\sqrt[3]{2}}{3+4\sqrt[3]{2}}$~\cite{LOSU}.
As general superpositions can be obtained through\cite{LOSU,KENNLINIE} $\ket{\Psi[z]}:=\ket{W}+z \ket{GHZ}$ with a complex number $z$. Here we are in the situation that all wave function coefficients are real.
\footnote{The same applies to the unique ground state of a hermitean Hamiltonian.} 
Consequently we have $\tau_3[\Psi[z]]=\tau_3[\Psi[z^*]]^*$. This implies the further symmetry $|\tau_3|[z]=|\tau_3|[z^*]$. This facilitates the search for the structure of the convex-roof since both the state $N_1$ and the line intersecting $\ket{N_1}$ with $\ket{GHZ}$ must lie in the real plane. 

To emphasize the ease to generalize the calculations we mention that 
for the n-qubit generalized states $\ket{GHZ_n}\propto \ket{\id_n} +\ket{\mathbb{0}_n}$ and $\ket{W_n}\propto \sum_{i=1}^n \ket{i_n}$ and the n-tangle $\tau_n^{(2(n-1))}$ of 
polynomial degree $2(n-1)$ 
this symmetry would be $n$-fold \cite{OS04,OS05,DoOs08,osterloh2010invariant,johansson2014classification}. Herein $\ket{\id_n}$/$\ket{\mathbb{0}_n}$ means that every of the $n$ qubits is in the state $\ket{1}$/$\ket{0}$, and $\ket{i_n}$ is the state with every qubit in the state $\ket{0}$ but the $i$-th qubit, which is in the state $\ket{1}$. 
For every n-tangle $\tau_n^{2m}$ where $n-1$ does not divide $m$ the effect of adding $\ket{W_n}$ is the same of adding an (orthogonal) product state; the result with a zero-polytope being shrunk to a single point is exactly solvable\cite{regula2016geometric,regula2016entanglement}

\subsection{The pure entangled states $N_i$} \label{sec:states-Ni}
One of the pure entangled states is the maximally entangled GHZ state\cite{LOSU,viehmann2012rescaling}. This is also confirmed through the convexification procedure of $(2,1)$ and $(1,1)$ decompositions which correctly points to the GHZ state. Remains to find the state in one of the three vertical wedges of the Bloch sphere. 
It is worth choosing another orthogonal basis for this purpose. The orthogonal decomposition chosen here is between the real one of the three pure states of the zero-polytope and its orthogonal partner. These are given by
\beqa
\ket{Z_3}&=&\sqrt{p_0}\ket{GHZ}-\sqrt{1-p_0}\ket{W}\\
\ket{Z_{3;\perp}}&=&\sqrt{1-p_0}\ket{GHZ}+\sqrt{p_0} \ket{W} 
\eeqa
with $p_0=\frac{4\sqrt[3]{2}}{3+4\sqrt[3]{2}}$~\cite{LOSU} (the calculations are shown in Appendix \ref{app:states-Ni}).
The angle of the new state is $\theta=1.99158=114.109^\circ$. This corresponds to $p_N=0.295762$. The four tetrahedra
are shown in grey color in Fig.~\ref{fig:OptimDecomp-GHZ-W} with green edges. 
Invoking the three-fold symmetry for this case leads to the three optimal $(3,1)$ decompositions with the states $\ket{N_i}$, $i=1,2,3$, as a tip.
We emphasize that the simplex $\ket{GHZ}-\ket{Z_1}-\ket{Z_2}$ is co-planar with $\ket{Z_1}-\ket{Z_2}-\ket{N_1}$; so are the simplices 
$\ket{Z_1}-\ket{N_1}-\ket{W}$ and $\ket{Z_1}-\ket{N_3}-\ket{W}$.
The rotational symmetry by multiples of $2\pi/3$ implies that this holds for all the corresponding faces.

Symmetry requirements would then lead to three lines along the meridians connecting $\ket{GHZ}$ with $\ket{N_i}$ as possible continuum of optimal states in $(2,1)$ decompositions.
We demonstrate in the following subsection that these are indeed optimal.
Fig~\ref{fig:OptimDecomp-GHZ-W} shows these $(2,1)$ decompositions as blue lines on the Bloch sphere surface.

\subsection{$(n_z,n_e)$-decompositions with $n_e>1$}\label{sec:decomps}

We consider here $(n_z,2)$ decompositions to demonstrate that in this case $(2,1)$ decompositions are indeed optimal in the direction of $\ket{GHZ}$ and
$(1,1)$ decompositions are optimal in the direction of $\ket{W}$. 

Even in the presence of zero-state locking it is not clear that there are no $(0,n_e)$ decompositions 
to fill up the Bloch sphere. In this case no zero-state state is visible from the density matrix from existing optimal decompositions. As a consequence, there have to exist also optimal $(n_z,n_e)$-decompositions.
Assume $(n_z,n_e)$-decompositions would exist.
In the specific case of rank two density matrices, we can limit ourselves with $(n_z,2)$-decompositions with $1\leq n_z\leq 2$ being subject to zero-state locking.
\begin{figure}[h!]
    \centering
    \includegraphics[width=.6\linewidth]{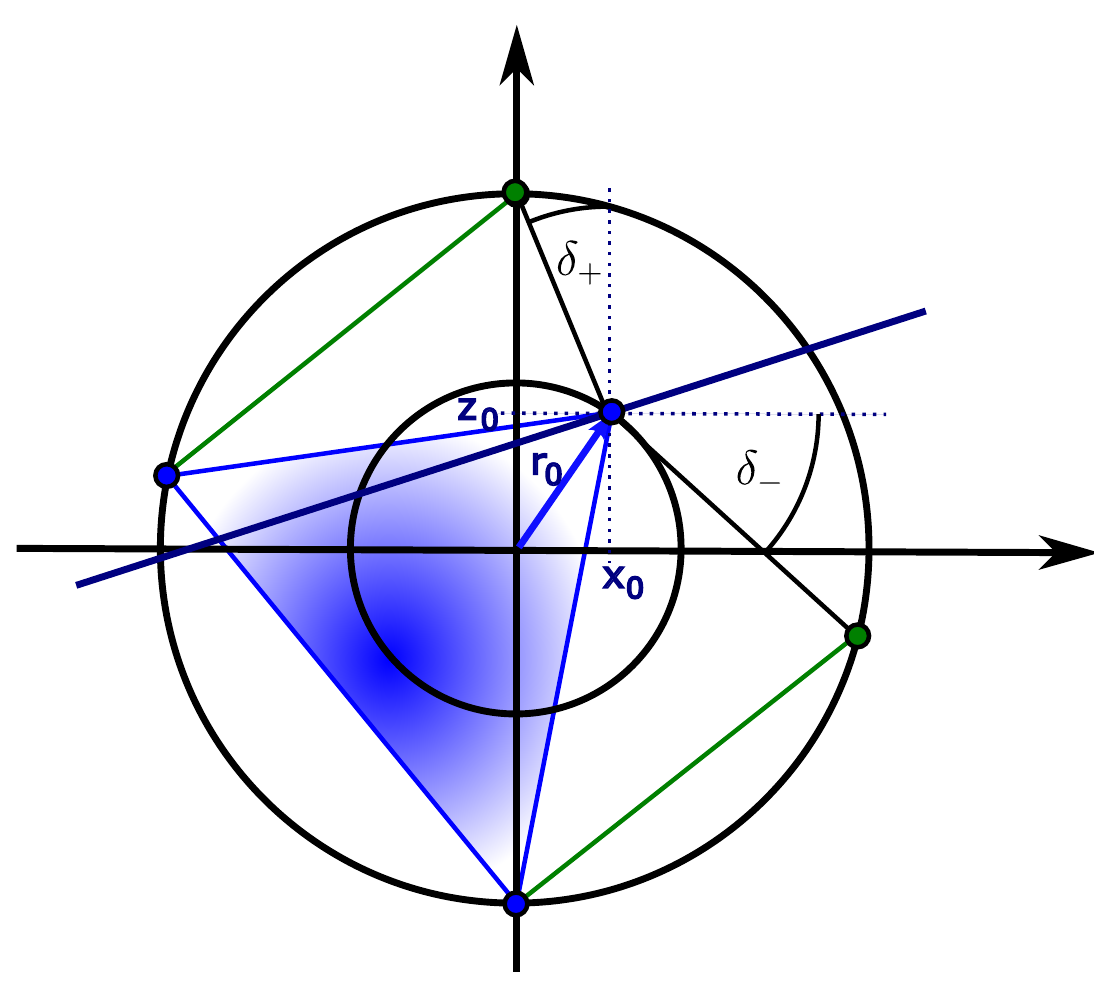}\hfill
     \includegraphics[width=.35\linewidth]{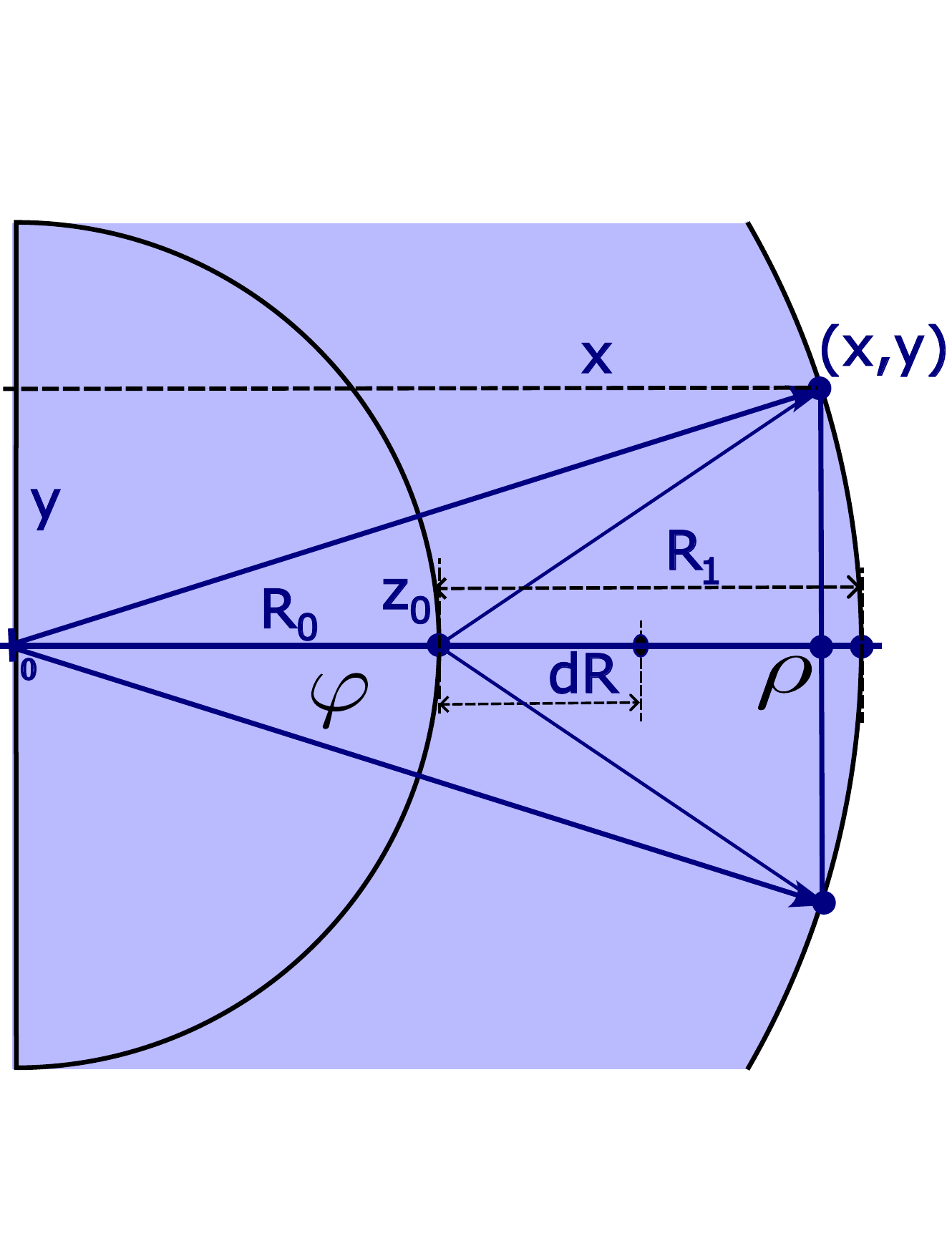}
   \caption{Left: The zero-polytope is shown (blue) in the Bloch sphere projection in $(x,z)$ plane, together with the sphere of radius $r_0$ given by the two complex conjugated zero states. The entangled pure states marking optimal decompositions are marked by green dots. The plane inclined by an angle $\delta_1$ is shown in purple color, cutting both spheres in circles of radii $R_0$ and $R=R_0+R_1$. The inclination angle is limited by the three-dimensional polytopes to the interval $[\delta_-,\delta_++\pi/2]$. $\delta_-$ is negative here. Right: The purple plane inclined by the angle $\delta$ within the Bloch sphere is shown together with part of the circles of radii $R_0$ and $R=R_0+R_1$. The entanglement of the density matrix $\rho$ decomposed of the two complex conjugated states located in $(x,\pm y)$ on this plane should be lower than the $(2,1)$ decomposition in consideration. For any $(2,2)$ decomposition being optimal for a state $\rho'$ located at distance $dR$ from the point $z_0$, there need to be states $\rho$ that have to be better decomposed by two corresponding states in $(x,\pm y)$.}
    \label{fig:decompositions}
\end{figure}

We assume that we have only real wavefunction coefficients to include the mirror symmetry of the Bloch sphere along the $x$-$z$ axis. In particular
the solutions to the zero polytope are constituted of complex conjugated pairs.
Let two such solutions be $z$ and $z^*$ with a real part $\Re z\stackrel{\wedge}{=}(x_0,0,z_0)$ in the Bloch-sphere which has a distance $r_0=\sqrt{x_0^2+z_0^2}$ to the origin (see left panel of Fig.~\ref{fig:decompositions}). Its polar angle is $\vartheta=\arctan(x_0/z_0)$. The normalized vector $\vec{n}=(\cos(\vartheta_1),\sin(\vartheta_1))$ describes the direction of the projected plane (purple line/plane in the left/right panel of Fig.~\ref{fig:decompositions}) onto what corresponds to the real plane in Hilbert space. This line $\vec{z}_0+\lambda~\vec{n}$ hits the Bloch sphere surface in the two intersection points 
\beqa
\lambda_{1/2}&=&-r_0 |\cos(\vartheta-\vartheta_1)|\pm \sqrt{1-r_0^2 \sin^2(\vartheta-\vartheta_1)}\; \\
&=& -R_0\pm R
\eeqa
where $R_0=r_0  |\cos(\vartheta-\vartheta_1)|$ is the radius of the inner and $R=\sqrt{1-r_0^2 \sin^2(\vartheta-\vartheta_1)}$ of the outer circle (see right panel of Fig.~\ref{fig:decompositions}).
According to Fig.~\ref{fig:decompositions}, at the distance $x(\delta \varphi)$ 
from the zero state(s) is the state $\rho$ of consideration. Defining $\Delta x(\delta \varphi)$ such that $x(\delta \varphi)+\Delta x(\delta \varphi)=R_1$, a straight forward calculation yields
\beq
\Matrix{c}{\Delta x(\delta\varphi)\\y(\delta\varphi)}=R\Matrix{c}{1-\cos\,\delta\varphi\\ \sin\,\delta\varphi} \approx R \, \Matrix{c}{\delta\varphi^2/2\\ \delta \varphi}
\eeq
In this plane the tangle will behave as 
\beq
\tau\approx\tau_0+ \frac{\tau"(0)}{2}\delta\varphi^2
\eeq
If $\tau"(0)>0$ there is a minimum of $\tau$ and the $(2,1)$ decomposition is clearly better. Hence we assume that 
$\tau"(0)<0$.
For the $(2,1)$ decomposition we obtain $\tau_{(2,1)}=\tau_0 (R_1-\Delta x(\delta \varphi))/R_1=\tau_0(1-\frac{R}{2R_1}\delta\varphi^2)$
and for the $(0,2)$ decomposition 
$\tau_{(0,2)}=\tau_0(1-\frac{|\tau"(0)|}{2 \tau_0}\delta\varphi^2)$.
Thus $\tau_{(2,1)}-\tau_{(0,2)}\approx (|\tau"(0)|-\tau_0\frac{R}{R_1})$.
This difference vanishes {\em iff} 
$|\tau"(0)|=\tau_0\frac{R}{R_1}$.
It is positive iff $\tau_{(0,2)}<\tau_{(2,1)}$. This leads to 
\beq\label{eq:2-1vs0-2}
-\frac{\tau"(0)}{\tau_0}>\frac{R}{R_1}\; .
\eeq
\begin{figure}[h!]
    \centering
    \includegraphics[width=1.\linewidth]{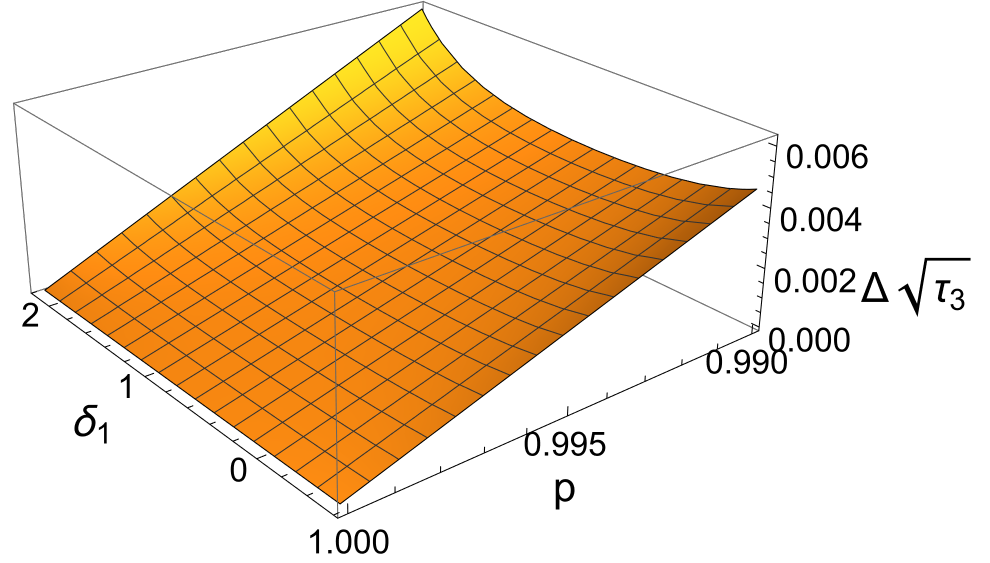}
   \caption{Difference $\Delta\tau:=\tau_{(0,2)}-\tau_{(2,1)}$ of the (0,2) and $(2,1)$ decompositions is shown for $\tau:=\sqrt{\tau_3}$ is shown depending on $\delta_1$ for $p$ close to $1$.}
    \label{fig:2-1vs0-2-decompositions}
\end{figure}
A remark is in order as the variable $\varphi$ is not the standard azimuthal angle of the Bloch sphere. It is to be taken in the plane 
described by the purple plane in Fig.~\ref{fig:decompositions}.

Here, the tangle is the square-root of the threetangle. 
With $\ket{GHZ}=(\ket{000}+\ket{111})/\sqrt{2}$ and $\ket{W}=(\ket{100}+\ket{010}+\ket{001})/\sqrt{3}$ 
being the poles of the Bloch sphere, the pure states' tangle is
\beqa
\sqrt{|\tau_3|}(p,\varphi)&=&\left| p^2+\frac{2^{7/2}}{3^{3/2}} \sqrt{p(1-p)^3}e^{i 3 \varphi}\right|^{1/2} 
\eeqa
\beqa
&=& \left[p^4+\frac{32 p^2}{\sqrt{54}}\sqrt{p(1-p)^3}\cos 3\varphi \right . \nonumber \\
& &\qquad\qquad\qquad\quad + \left .\frac{16^2}{54}p(1-p)^3\right]^{1/4}
\eeqa
In the characteristic curves in Ref.~\cite{LOSU} (taking the square root) it is seen that the strongest $\varphi$ dependence 
originates around $p_0$. At $p_0$ the variable varied is indeed the azimuthal angle of the Bloch sphere. We obtain
\beq
\sqrt{|\tau_3|}(p_0,\delta\varphi)=\sqrt{2}p_0\left(1-\frac{3^2}{4^2} \delta\varphi^2\right)\; .
\eeq
We have $-\sqrt{\tau_3}"(p_0,0)/\sqrt{\tau_3}(p_0,0)=\frac{9}{8}<2$ since $R=2 R_1$ for this case.
In Fig.~\ref{fig:2-1vs0-2-decompositions} the behavior is shown for all admissible angles $\delta_1$. 
Hence, the $(2,1)$ decomposition is optimal.
It is interesting that the characteristic lines follow the  relative maxima with respect to $\varphi$ of the corresponding tangle of the type $(n_0,1)$. Whether this is true in general could be easily checked for more general states of rank two.
This is certainly not the case for the further connecting lines with respect to $p$, as we will see in what follows.

\subsection{Optimal $(2,1)$ decompositions in between the states $\ket{N_i}$} \label{sec:states-Mi}
In order to get a glimpse of what is happening in between the lower tetrahedra, we briefly want to consider the case in which an orthogonal decomposition of the density matrix lies within this simplex of $(2,1)$ decompositions. For symmetry reasons, this is the case for the two pure states lying respectively on the vectors corresponding to $\ket{Z_1}+\ket{Z_2}$ or to $\ket{Z_3}+\ket{W}$.
The two cases lead to the same result (for the calculations see Appendix \ref{app:states-Mi}). 
The angle of the state $\ket{M_3}$ is 
$\theta=2.25566=129.24^\circ$. The three points corresponding to the states $\ket{M_i}$ for $i=1,2,3$ are marked in Fig.~\ref{fig:OptimDecomp-GHZ-W} with white dots. 

For the lines on the Bloch sphere, we rely on a direct comparison of $(2,1)$ with $(1,1)$ decompositions. We therefore chose the connecting line between $\ket{N_3}$ and $\ket{M_3}$; the complete solution is obtained by applying the symmetries and the details are described in Appendix~\ref{app:states-Mi}. 
This leads to the lower blue curves shown in Fig.~\ref{fig:OptimDecomp-GHZ-W} which lie on a circle with distance to the center of the Bloch sphere being $d_M=0.0711148$.
That the states form a circle is confirmed by calculating the derivative with respect to its angle
and showing its orthogonality to the original vector (not shown here).
The numerical error scales with $\Delta \phi^2$ as expected. The normal direction of the circle is $\vec{n}[M]=(0.57589,0,-0.81753)=:(c_1,0,c_3)$. Using the symmetry, we get the three normal vectors given by dashed lines in blue to the blue points on the sphere and for the polar grand circles as well.
That the $(2,1)$ decompositions are optimal can be independently checked looking at the tangle of the rank-two density matrix mixing the state $\ket{M_3}$ with its orthogonal complement $\ket{M_3^\perp}$. It indeed needs the convexification of $\ket{M_3}$ and the two zero-states $\ket{Z_3}$ and $\ket{W}$, a $(2,1)$ decomposition. For the states in consideration, equation \eqref{eq:2-1vs0-2} is falsified with 
\beq
2.55929=\frac{R}{R_1}\geq -\frac{\sqrt{\tau_3}"[M_3]}{\sqrt{\tau_3}[M_3]}=0.849718
\eeq
where the definitions of $R$, $R_1$ are following Fig.~\ref{fig:2-1vs0-2-decompositions} and $\tau"=\partial^2_\vartheta \tau$ in this case ($\tau$ is used for $\sqrt{\tau_3}$ for the sake of readability).
Hence, the $(2,1)$ decomposition is optimal; this applies to the whole line interconnecting the states $\ket{N_i}$, $i=1, 2, 3$.
It is
\beqa
\vec{r}\cdot\vec{n}[M]&:=&\Matrix{c}{\sin \vartheta \cos \varphi \\
\sin \vartheta \sin \varphi \\
\cos \vartheta}\cdot \vec{n}[M]\nonumber \\
&=& c_1 \sin \vartheta \cos \varphi + c_3 \cos \vartheta \\
\cos \vartheta_M[\varphi] &=& c_3 d_M \nonumber \\
&-& \!\!\!\!\!\frac{\sqrt{c_3^2 d_M^2+ (c_1^2 \cos^2 \vartheta - d_M^2)(c_3^2 + c_1^2 \cos^2 \varphi)}}{c_3^2 + c_1^2 \cos^2 \varphi} 
\eeqa

We comment on what happens {\em iff} Eq.~\eqref{eq:2-1vs0-2} was satisfied. 
Having the optimal $(2,2)$ decompositions given by a convexification procedure for two inner states, $\ket{M_{(0,2);1}}$ and $\ket{M_{(0,2);2}}$, in the plane $P$ given by the three states in consideration (here these are $\ket{W}$, $\ket{Z_3}$, and $\ket{M_3(\varphi)}$), the tangle would start being strictly convex from here. I do not know of an exact solution in order to establish the missing $(1,1)$ decompositions. It is this non-convex part that will render a solution to the convex-roof problem in general NP-hard.
However, this would apply only to the regions in the vicinity of the Bloch sphere surface. Therefore its tangle will be in the range $[\min\{\tau[M_{(0,2);1}],\tau[M_{(0,2);2}]\},\max_C\{\tau[\psi_C]\}]$,
where $C$ is the curve of the Bloch sphere surface with the plane $P$ in between the both states $\ket{M_{(0,2);i}}$, $i=1,2$.

For the remaining parts of the Bloch sphere there is only a single visible state from the zero polytope left.
This leads to $(1,1)$ decomposition being optimal. They are depicted in Fig.~\ref{fig:OptimDecomp-GHZ-W} by green lines and red points.

\section{Conclusions}
The mixture of GHZ and W state has been elaborated for all the Bloch sphere, obtaining a pattern on the surface that is characteristic for these two states in the mixture. We give a proof of the zero-state locking where the finding is based on. We also derive an inequality which decides whether $(n_z,n_e)$ decompositions for $n_e>1$ are optimal and find them  not optimal for the present case. 
For the zero-state locking it is essential to look at certain roots which render the tangle to scale linear with the density matrix. This is the square root in case of the threetangle.

The Bloch sphere is divided into the zero polytope and four polytopes onto each face of the zero polytope with corresponding four pure states. The faces of two neighboring such polytopes are strictly planar. There are pure states that belong to $(2,1)$ decompositions moving on three polar grand circles 
and three circles which possess a small distance $0.0711148$ to the Bloch sphere center. Their normal vector is $(0.57589 \cos \phi_i,0,-0.81753 \sin \phi_i)$, with $\phi_i=i 2\pi/3$. The remaining decompositions to fill the Bloch sphere are of the type $(1,1)$.
This distribution of the optimal decomposition inter-connects the polytopes of various dimensions. 
For SL invariant tangles there is a transformation rule that applies in particular to the optimal decomposition states\cite{viehmann2012rescaling}. 
Applying these rules yields the solution for the convex-roof in the transformed density matrix. For the general case the characteristic features of the convex roof will modify, but the structure will stay unaltered, as reported also in Ref.~\cite{neveling2024threetangle}. It would be interesting to what extent these findings are universal. The paths of the pure entangled state of $(2,1)$ decompositions here moves on circles, but this will not apply to the general case. It is intriguing to look for examples where the inequality points towards $(0,2)$ decompositions in parts of the Bloch sphere is satisfied. Future investigations, clarifying these questions, will be needed.
This work describes the way how optimal decompositions of rank two mixtures behave. It will be a challenge to understand their behavior for higher ranks. This can also become relevant for various quantum technologies where an n-tangle is needed. Future work should establish the specific role of genuine SL-invariant entanglement in this respect.

{\em Acknowledgements --}
I acknowledge the TII and Luigi Amico for supporting this research.
\\

\appendix

\section{Detecting genuine three-partite entanglement via the threetangle}\label{app:threetangle}

We will consider $\sqrt{|\tau_3|}$ as entanglement measure, where
the threetangle $|\tau_3|$ has been defined as\cite{Coffman00} 
(see also in Refs.~\cite{Wong2001,VerstraeteDM03NormalForms,OS04})
\nbeqa
\tau_3 &=& d_1 - 2d_2 + 4d_3  \\
  d_1&=& \psi^2_{000}\psi^2_{111} + \psi^2_{001}\psi^2_{110} + \psi^2_{010}\psi^2_{101}+ \psi^2_{100}\psi^2_{011} \\
  d_2&=& \psi_{000}\psi_{111}\psi_{011}\psi_{100} + \psi_{000}\psi_{111}\psi_{101}\psi_{010}\\ 
    &&+ \psi_{000}\psi_{111}\psi_{110}\psi_{001} + \psi_{011}\psi_{100}\psi_{101}\psi_{010}\\
    &&+ \psi_{011}\psi_{100}\psi_{110}\psi_{001} + \psi_{101}\psi_{010}\psi_{110}\psi_{001}\\
  d_3&=& \psi_{000}\psi_{110}\psi_{101}\psi_{011} + \psi_{111}\psi_{001}\psi_{010}\psi_{100}\ \ ,
\neeqa
and coincides with the three-qubit hyperdeterminant\cite{Cayley,Miyake02}. 
It detects states from the only genuine SL-entangled GHZ-class here.
$W$-states are not detected; they are instead detected as entangled by the pairwise concurrence which is distributed along all possible pairs in the state. It is therefore not bipartite
and is called {\em genuinely multipartite entangled} in the literature. Occams razor would forbid giving more than one name to one and the same quantity: here non-bipartiteness.
However, it is only genuinely pairwise SL-entangled.

\section{General properties of optimal decompositions for real states of rank two}\label{app:optimal-decompositions}

For the number $n_{\rm opt}$ of pure states contained in any optimal decomposition holds ${\rm rank} [\rho]
\leq n_{\rm opt} \leq ({\rm rank} [\rho])^2$~\cite{caratheodory1911variabilitatsbereich,Uhlmann98}.
Thus, it is sufficient to look for up to $4$ such pure states that generically form up to three-dimensional simplices.
We want to emphasize that more than the maximal $4$ states may be in an optimal decomposition; it is given by the convex polytope made out of these points. The tangle will behave linearly in these optimal polytopes.
In this case, since every sub-partition of an optimal decomposition is itself optimal, each $4$ states out of that decomposition are optimal as well.

Since the states are composed of real elements,
every tangle $\tau$ has the property that $\tau(z)\equiv \tau(z^*)$ where $\ket{\psi}\propto\ket{\psi_1}+z \ket{\psi_2}$ is a representation that corresponds to a vector onto the Bloch-sphere with the parametrization\cite{osterloh2016exact}
\beq
z \mathrel{\widehat{=}}\vec{n}=\Matrix{c}{
x \\
y \\
z
}=\Matrix{c}{
2\sqrt{p(1-p)}\cos\phi \\
2\sqrt{p(1-p)}\sin\phi \\
2p-1
}
\eeq
where $\ket{GHZ}$ and $\ket{W}$ are the two anti-podes on the z-axis.
Real superpositions lie in the $x$-$z$-plane.

We will be mainly interest in a) where the line $\overline{\vec{n}_0\vec{n}_1}$ cuts the $z$-axis,  b) which are the weights of the corresponding entangled state(s), and c) where it crosses the Bloch sphere,
that is the corresponding pure states.
Whereas the answer to a) is given by the probability
\beq
P=\frac{p_0\sqrt{p_1(1-p_1)}\cos\phi_1+p_1\sqrt{p_0(1-p_0)}\cos\phi_0}{p_0\sqrt{p_0(1-p_0)}\cos\phi_0+p_1\sqrt{p_1(1-p_1)}\cos\phi_1}
\eeq
such that the point is located in $\vec{n}_P=(2P-1)\vec{e}_z$,
the answer to b) is given by 
\beq
\lambda=\frac{m_1}{m_0}=\frac{\sqrt{p_0(1-p_0)}\cos\phi_0}{\sqrt{p_1(1-p_1)}\cos\phi_1}
\eeq
where $m_i$ is the weight of the corresponding state.
For the mainly interesting case of $\phi_1=0$ and $p$ being the probability in the density matrix in consideration, we obtain as a solution to c)
\begin{widetext}
\beq
P_\pm=p_0+\frac{p-p_0}{2}\cdot\frac{(p-p_0)(1-2 p_0)+2 p_0(1-p_0)\cos^2\phi_0\pm\sqrt{(p_0-p)^2+4 p p_0(1-p_0)(1-p)\cos^2\phi_0})}{(p-p_0)^2+ p_0(1-p_0) \cos^2 \phi_0}
\eeq
\end{widetext}
The resulting tangle is a convex combination of the tangle values of the respective pure states.
\begin{figure}[h!]
    \centering
    \includegraphics[width=\linewidth]{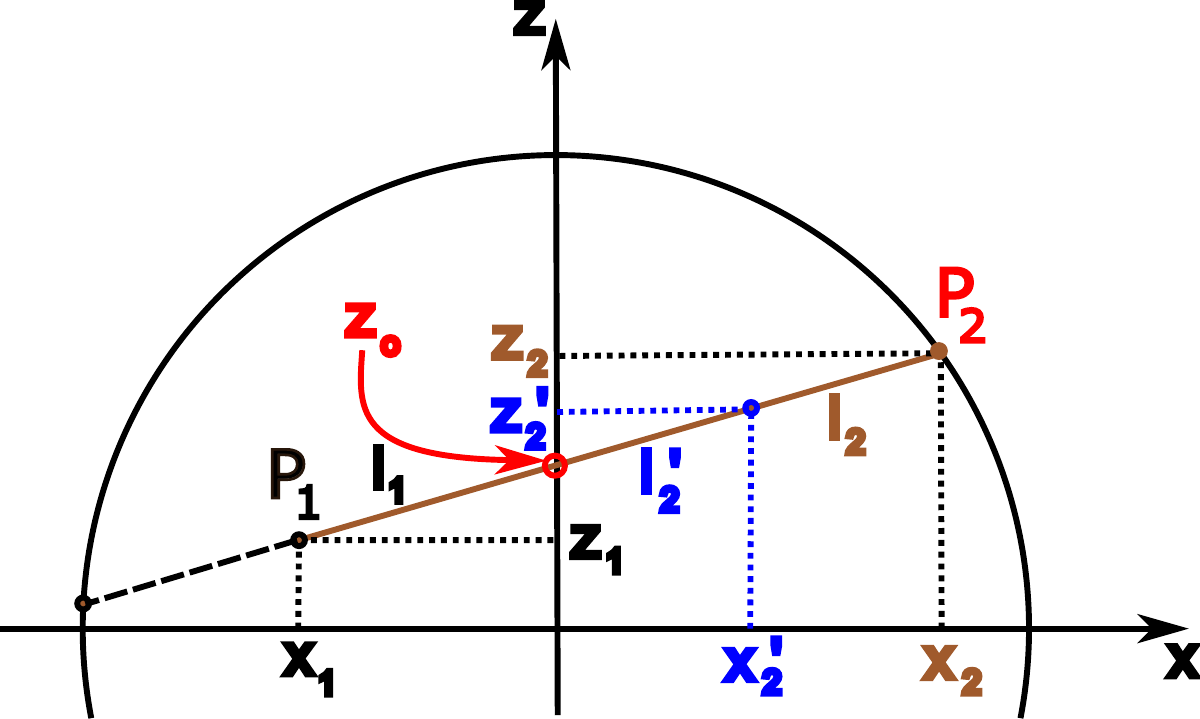}
   \caption{States within the Bloch sphere. For a given line $\overline{P_1P_2}$ all triangles $P_0P_1Z_1$, $Z_0P_2Z_2$, and $Z_0P_2'Z_2'$, are similar. $P_i=(x_i,z_i)^t$, $Z_i=(0,z_i)^t$; $t$ for transposed. Ratios of their respective lengths are determined by the theorem of intersecting lines.}
    \label{fig:intersectinglines-optimDecomp}
\end{figure}
This strictly linear behavior inside all the simplices of an optimal decomposition indirectly tells about whether there needs to be one or more states in a decomposition for being optimal:
the continuation of the respective decomposition type has to lead to a convex behavior in the tangle value; where this condition is not satisfied a (linear) convexification is needed. This holds in particular if the corresponding optimal polytope changes dimension. Here, this means to a dimensionality of at most three.


\subsection{$(1,1)$-decompositions}

Here, we collect the formulae obtained for $(1,1)$-decompositions of a pure state at z-coordinate $2p_1-1$, where $p_1$ corresponds to its probability of the two states $\ket{\psi_i}$; $i\in\{N,S\}$. This corresponds to two pure states in the left panel of Fig.~\ref{fig:intersectinglines-optimDecomp}.
We find
\beqa
l_1(p_1,p)&=&2\sqrt{(p-p_1)^2+p_1(p_1-1)}\\
l_2(p_1,p)&=&\frac{4p(1-p)}{l_1(p_1,p)}\\
m_2(p_1,p)&=&\frac{l_1(p_1,p)}{l_1(p_1,p)+l_2(p_1,p)}\\
p_2(p_1,p)&=&\frac{4p^2(1-p_1)}{l_1(p_1,p)^2}\\
&=&(1-p_1)\frac{p}{1-p}\cdot\frac{l_2(p_1,p)}{l_1(p_1,p)}\\
p(p1,p2)&=&p_1+\frac{(p_2-p_1)\sqrt{p_1(1-p_1)}}{\sqrt{p_1(1-p_1)}+\sqrt{p_2(1-p_2)}}
\eeqa
Next we want to examine whether for a given $(1,1)$ decomposition with $p_1$, $p_2$, and $p$ for a given pure state it is convenient to split the pure state into two, having a $(1,2)$ decomposition. We want to emphasize that the density matrix will be a point on the line in general: this line is kept fixed.
It can be seen that the ratios $x_1/(2|p_1-p|)=x_2/(2|p_2-p|)=x'_2/(2|p'_2-p|)$ are equal with $y'_2=\pm\sqrt{4p'_2(1-p'_2)-{x'_2}^2}$ (see left panel in Fig.~\ref{fig:intersectinglines-optimDecomp}).
For the new $p'_2$ of this new state at phase $\phi_2$ the calculation gives
\begin{widetext}
\beq
p'_{2}(p_1,p;\phi_2)=p_1+(p-p_1)\frac{2p_1(1-p_1)-(p-p_1)(2p_1-1)\cos^2 \phi_2+\cos \phi_2\sqrt{4p(1-p)p_1(1-p_1)+(p-p_1)^2 \cos^2 \phi_2}}{2(p_1(1-p_1)+(p-p_1)^2\cos^2 \phi_2)}
\eeq
\end{widetext}

\begin{figure}[h!]
    \centering    
    \includegraphics[width=\linewidth]{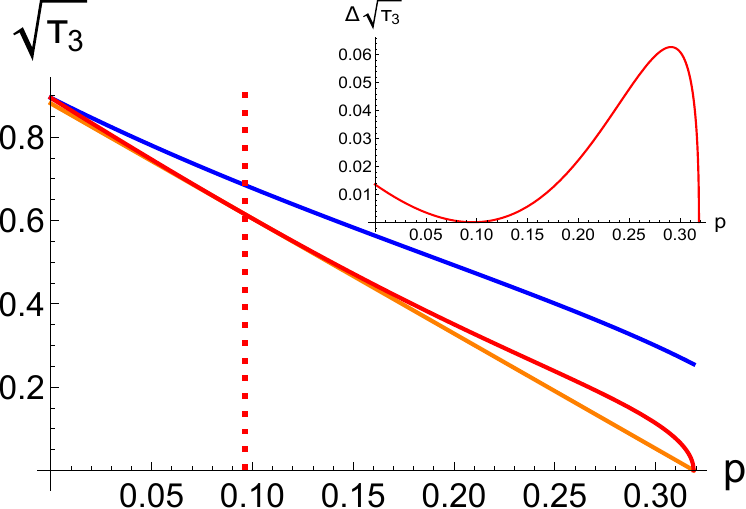}
   \caption{Convexification procedure of $\ket{Z3}$ mixed with the corresponding orthogonal state.
   The curves are made convex with a line touching at $p_c=0.0964142$, as is highlighted in the inset, were we plot the difference in $\sqrt{\tau_3}$ of the $(1,1)$ decompositions with $\ket{Z_3}$ and the convex line. The $(2,1)$ decompositions with $\ket{Z_3}$ and $\ket{W}$ are shown in blue above both curves.}
    \label{fig:Z3-convexification}
\end{figure}

\subsection{$(2,2)$-decompositions}

In the following we will assume that two complex conjugated pairs form a $(2,2)$ decomposition, corresponding to a superposition of two states at $p_i$ and angles $\pm \phi_i$, $i\in\{1,2\}$, such that both states come to lie inside the Bloch sphere projection.
We find
\begin{widetext}
\beqa
m_2(p_1,\phi_1;p_2,\phi_2)&=&\frac{\sqrt{p_1(1-p_1)\cos^2(\phi_1)}}{\sqrt{p_1(1-p_1)\cos^2(\phi_1)}+\sqrt{p_2(1-p_2)\cos^2(\phi_2)}}\\
&=&
\frac{l_1(p_1,\phi_1;p_2,\phi_2;p)}{l_1(p_1,\phi_1;p_2,\phi_2;p)+l_2(p_1,\phi_1;p_2,\phi_2;p)}\\
m_1(p_1,\phi_1;p_2,\phi_2)&=& 1-m_2(p_1,\phi_1;p_2,\phi_2)\\
p(p_1,\phi_1;p_2,\phi_2)&=&p_1+(p_2-p_1)\frac{\sqrt{p_1(1-p_1)\cos^2(\phi_1)}}{\sqrt{p_1(1-p_1)\cos^2(\phi_1)}+\sqrt{p_2(1-p_2)\cos^2(\phi_2)}}\\ &=&\frac{p_2\sqrt{p_1(1-p_1)\cos^2(\phi_1)}+p_1\sqrt{p_2(1-p_2)\cos^2(\phi_2)}}{\sqrt{p_1(1-p_1)\cos^2(\phi_1)}+\sqrt{p_2(1-p_2)\cos^2(\phi_2)}}\\
&=& p_1 m_1(p_1,\phi_1;p_2,\phi_2)+p_2 m_2(p_1,\phi_1;p_2,\phi_2)\\
l_1(p_1,\phi_1;p_2,\phi_2;p)&=&2\sqrt{(p_1-p)^2+p_1(1-p_1)\cos^2 \phi_1}\\
l_2(p_1,\phi_1;p_2,\phi_2;p)&=&2\sqrt{(p_2-p)^2+p_2(1-p_2)\cos^2 \phi_2}
\eeqa
\end{widetext}
From this it is seen that the weights of the states are symmetric under exchanging $p_i\to 1-p_i$ separately for $i\in\{1,2\}$.
Setting $\phi_2=0$, we are in $(2,1)$ decompositions and the result is
\begin{widetext}
\beq
p_2(p_1,\phi_1;p)=p_1+(p-p_1)\frac{(p_1-p)(2p_1-1)+2p_1(1-p_1)\cos^2 \phi_1+\sqrt{(p_1-p)^2+4p(1-p)p_1(1-p_1)\cos^2 \phi_1}}{2((p_1-p)^2+p_1(1-p_1)\cos^2 \phi_1)}
\eeq
\end{widetext}
Next, for the evaluation  of $(2,2)$ from $(2,1)$ decompositions for given $p_1$, $\phi_1$, and $p$ the ratios of the following quantities are kept fixed (see left panel of Fig.~\ref{fig:intersectinglines-optimDecomp}): $x_1/(2|p_1-p|)=x_2/(2|p_2-p|)=x'_2/(2|p'_2-p|)$ with $y'_2=\pm\sqrt{4p'_2(1-p'_2)-{x'}_2^2}$. These have to be substituted in $m_2(p_1,\phi_1;p'_2,\phi'_2)$ where $x'_2=2\sqrt{p'_2(1-p'_2)}\cos \phi'_2$ to establish the weights.

\begin{figure}[h!]
    \centering    
    \includegraphics[width=\linewidth]{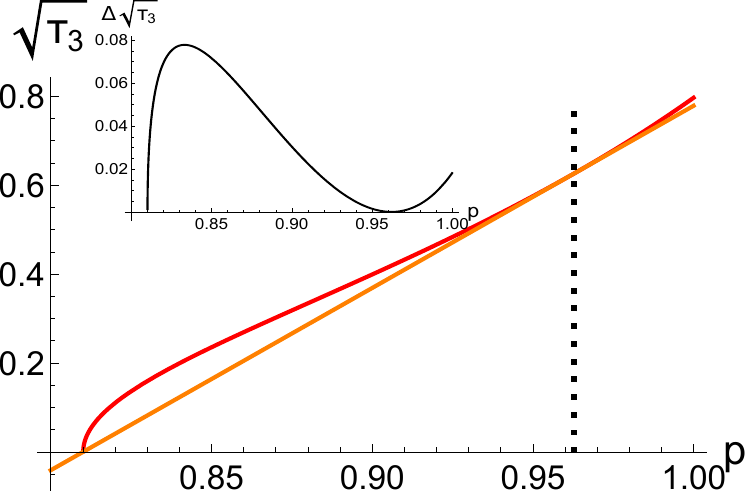}
   \caption{Convexification procedure of the two corresponding pure states lying on the line given by the vector $\ket{Z_1}+\ket{Z_2}$ in a mixed state.
   The curves are made convex with a line touching at $p_c=0.962243$, as is highlighted in the inset, were we plot the difference in $\sqrt{\tau_3}$ of the $(1,1)$ decompositions with $\ket{Z_3}$ and the convex line. The $(2,1)$ decompositions with $\ket{Z_3}$ and $\ket{W}$ are above both curves; they are not shown here.}
    \label{fig:Z1+Z2-convexification}
\end{figure}

\section{The pure entangled states $\ket{N_i}$}
\label{app:states-Ni}

We analyze the Bloch sphere of the rank-two mixed state
\beq
\rho[p]=p\ketbra{GHZ} + (1-p) \ketbra{W}
\eeq
where $\ket{GHZ}=(\ket{111}+\ket{000})/\sqrt{2}$ and $\ket{W}=(\ket{100}+\ket{010}+\ket{001})/\sqrt{3}$.
A three-fold rotational symmetry due to cyclic permutations of the qubits exist for both states. Therefore,
it is enough to perform the classification in a wedge of the Bloch sphere given by the azimuthal angle $\phi\in [-\pi/3 , \pi/3]$ only. The rest of the sphere is obtained by symmetry requirements.
\\
This analysis verifies a known feature, namely that the pure entangled state is the maximally entangled GHZ state\cite{LOSU,viehmann2012rescaling}. This former result is indeed confirmed through the convexification procedure of $(2,1)$ and $(1,1)$ decompositions which correctly points to the GHZ state. Remains to find the state in one of the three vertical wedges of the Bloch sphere. 
One possibility is to choose another orthogonal basis for the same Bloch sphere. The one orthogonal decomposition chosen here is the real one of the three pure states of the zero-polytope together with its orthogonal partner. 
\begin{figure}[h!]
    \centering    
    \includegraphics[width=0.48\linewidth]{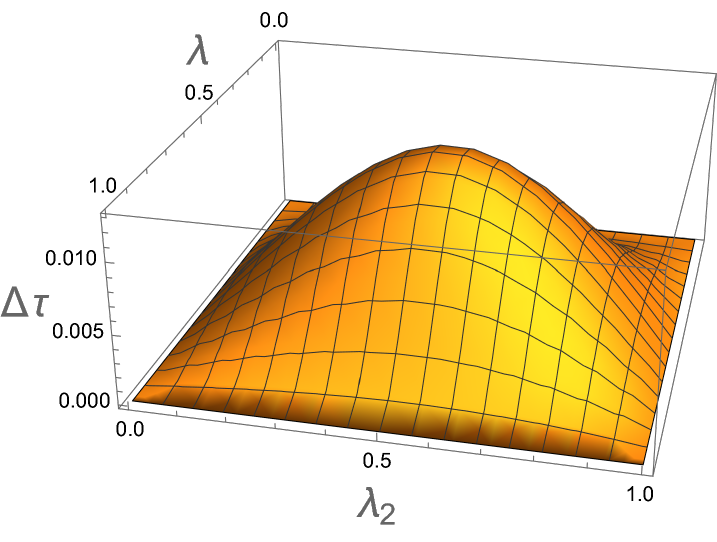}\hfill
    \includegraphics[width=0.48\linewidth]{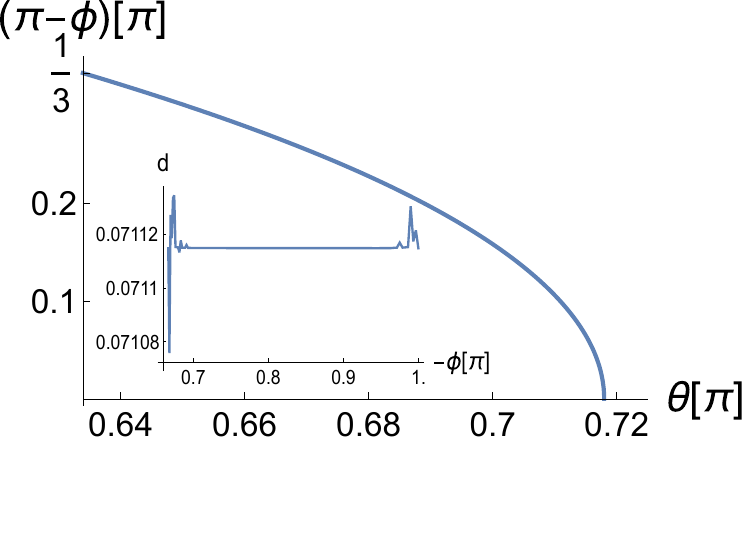}
   \caption{Left: Difference of $\sqrt{\tau_3}$ of $(1,1)$ and $(2,1)$ decompositions of $\ket{W}$ and $\ket{Z_3}$ for states connecting $\ket{N_3}$ and $\ket{M_3}$. Right: Positions of the entangled pure state in the $(2,1)$ decompositions of $\ket{W}$ and $\ket{Z_3}$ for states connecting $\ket{N_3}$ and $\ket{M_3}$. The inset shows the calculated distance of the plane going through three calculated states to the center of the Bloch sphere.}
    \label{fig:diff-2-3-decomps}
\end{figure}
These are given by
\beqa
\ket{Z_3}&=&\sqrt{p_0}\ket{GHZ}-\sqrt{1-p_0}\ket{W}\\
\ket{Z_{3;\perp}}&=&\sqrt{1-p_0}\ket{GHZ}+\sqrt{p_0} \ket{W} 
\eeqa
with $p_0=\frac{4\sqrt[3]{2}}{3+4\sqrt[3]{2}}$~\cite{LOSU}.
The relevant charcteristic curves for $(1,1)$ decompositions with the W state and $(2,1)$ decompositions with $(\ket{Z_1}+\ket{Z_2})/\sqrt{2}$ are shown in Fig.~\ref{fig:Z3-convexification} together with their convexification. 
It shows that at $p_c=0.0964142$ the line is a proper convexification, starting from zero and touching the curve for $(2,1)$ decompositions; the $(1,1)$ decompositions are shown in blue. The corresponding pure state is located in $p_2(p_1,\phi_1,p_c)=0.00673174$.
Since $p_c<1/2$, this corresponds to an angle $\theta_1=\pi-\arccos{(2p_c-1)}=0.164279=9.4125^\circ$, which adds up to the angle of $\ket{Z_{3;\perp}}$, which is 
$\theta_0=1.8273=104.697^\circ$. The angle of the new state is hence calculated to be $\theta=\theta_0+\theta_1=1.99158=114.109^\circ$. The four tetrahedra
are shown in grey in Fig.~\ref{fig:OptimDecomp-GHZ-W} with green edges. 
This is confirmed from the convexification procedure with the $(1,1)$ decompositions (with $\ket{W}$ here) for eigenstates being situated at an angle $\theta=140^\circ$. 
Invoking the three-fold symmetry for this case leads to the three optimal $(3,1)$ decompositions with the states $\ket{N_i}$, $i=1,2,3$, as a tip.
This shows as well that in this case $(2,1)$ decompositions are optimal in the direction of $\ket{GHZ}$ and
$(1,1)$ decompositions are optimal in the direction of $\ket{W}$. 
Fig~\ref{fig:OptimDecomp-GHZ-W} shows these $(2,1)$ decompositions as blue lines on the Bloch sphere surface.
In~\cite{neveling2024threetangle}, much effort has been given concerning the distribution of optimal decompositions. These findings hold for even more general SL-invariant tangles $\tau$ than the threetangle, detected by $\sqrt{\tau_3}$, and the observations here agree with it.
We emphasize that the simplices of $\ket{GHZ}-\ket{Z_1}-\ket{Z_2}$ is co-planar with $\ket{Z_1}-\ket{Z_2}-\ket{N_1}$; so are the simplices 
$\ket{Z_1}-\ket{N_1}-\ket{W}$ and $\ket{Z_1}-\ket{N_3}-\ket{W}$.
\section{Optimal $(2,1)$ decompositions in between the states $\ket{N_i}$ -- the states $\ket{M_i}$} \label{app:states-Mi}

In order to get a glimpse of what is happening in between the lower tetrahedra, we briefly want to see the case in which two orthogonal pure states lie within a simplex of $(2,1)$ decompositions. For symmetry reasons (mirror symmetry due to the reality of the wave function coefficients), this is the case for the two pure states lying respectively in the piercing points of the straight lines connecting the density matrices $\ket{Z_1}+\ket{Z_2}$ or $\ket{Z_3}+\ket{W}$ with the center of the Bloch sphere.
Both cases lead to the same result, and we will
describe explicitly the first case of $\ket{Z_1}+\ket{Z_2}$ in the Appendix \ref{app:states-Mi}.
The corresponding characteristic curves can be viewed in Fig.~\ref{fig:Z1+Z2-convexification}.
It is seen that a convexification leads here to a
$p_c=0.962243$ which corresponds to $p_2(p_0,\pi,p_c)=0.989858$ for the corresponding pure state on the Bloch sphere surface.
This leads to an angle $\theta_1=0.201758=11.5599^\circ$.
As before we have to consider that the two eigenstates lie at an angle 
$\theta_0=2.0539=117.68^\circ$.
So, the angle of the state $\ket{M_3}$ is 
$\theta=\theta_0+\theta_1=2.25566=129.24^\circ$. The three states $\ket{M_i}$ for $i=1,2,3$ are marked in the right panel in Fig.~\ref{fig:OptimDecomp-GHZ-W} with white dots.
For the further points in the Bloch sphere, we rely on a direct comparison of $(2,1)$ with $(1,1)$ decompositions. We therefore chose the connecting line between $\ket{N_3}$ and $\ket{M_3}$; the complete solution is obtained by applying the three-fold symmetry. It is important noticing that, at first sight, the effective tangle of $(2,1)$ and $(1,1)$ decompositions yield the same result.
That this is not so is seen by looking at the difference between $(1,1)$ and $(2,1)$ decompositions, shown in the left panel of Fig.~\ref{fig:diff-2-3-decomps}.
$\lambda$ refers to the parameter convexly connecting
$\ket{N_3}$ and $\ket{M_3}$, which gives the density matrix considered here, whereas $\lambda_2$ is the corresponding parameter for $\ket{W}$ and $\ket{Z_3}$, hence of the mixed state of both zero-states.
It is seen that the states under consideration get optimised by choosing the $(2,1)$ decomposition.
We optimize the tangle with respect to $\lambda_2$. The pure states are located in the positions indicated in the right panel of Fig.~\ref{fig:diff-2-3-decomps}. 
This leads to the lower blue curves shown in the right panel of Fig.~\ref{fig:intersectinglines-optimDecomp} which lie on a circle that is not a grand circle as the ones before; instead it has a distance to the center of the Bloch sphere of $0.0711148$.
This is calculated by planes going through three calculated states to the center of the Bloch sphere. In addition to the states $N_i$ and $M_j$ in between we minimize the tangle on a finite but arbitrary grating. We modified the distance the points should have on the grating. The result did not depend on this choice. The result equals this for the two states $N_i$ and the state $M_j$ in between.
There are numerical errors (due to a finite grating) to this, but they are small (smaller than $4$\textperthousand) and only occur in the border regions, where the difference to the $(1,1)$ decompositions tends to vanish (see inset of the right panel of Fig.~\ref{fig:diff-2-3-decomps}).
That the states form a circle is confirmed by calculating the derivative with respect to the angle
and showing its orthogonality to the original vector (not shown here).
The numerical error scales with $\Delta \phi^2$ as expected. The normal direction of the circle is $(0.57589,0,-0.81753)$. Using the symmetry, we get the three points of normal vectors together with a dashed line in blue. The same is done for the azimuthal grand circles.\\
We leave a note here on $(2,2)$ decompositions. The tangle is here not in a maximum but it is of course analytic. It has the Taylor expansion
\beq
\tau \approx \tau_0 + \tau'(0) \delta\varphi+ \frac{\tau"(0)}{2} \delta\varphi^2
\eeq
for the respective variables in the plane through the Bloch sphere as seen in Fig.~\ref{fig:decompositions}. 

If $(0,2)$ decompositions lead to lower results than the $(2,1)$ decomposition, so will do symmetric $(0,2)$ decompositions.
For symmetric $(0,2)$ decompositions we obtain again
$\tau_{(0,2)}=\tau_0+\frac{\tau"(0)}{2}\delta\varphi^2$
and the $(2,1)$ decomposition has again the form $\tau_{(2,1)}=\tau_0(1-\frac{R}{2 R_1}\delta\varphi^2)$.
Therefore, we get the same condition Eq.~\eqref{eq:2-1vs0-2} for the optimality of the $(0,2)$ decompositions.
This condition is falsified for the state $\ket{M_3}$,
with the ratio $\frac{\tau"(0)}{\tau_0}=-0.849718$ and the corresponding $R/(2R_1)=0.997468$. Therefore, the $(2,1)$ decompositions are optimal in this case. Indeed, looking at the density matrix
of the states $M_3$ and its orthogonal counterpart $M_3^\perp$ 
\beq
\rho_{M_3}=p\ketbra{M_3}+(1-p)\ketbra{M_3^\perp}
\eeq
the state $\ket{M_3}$ itself is the convexification point. 
In the same way, all the states on the circle $M(\varphi)$ are in optimal $(2,1)$ decompositions.
Symmetry extends these results to all $M_i$.

In general the density matrix where the convexification is needed would mark the border of the $(2,2)$ polytope. To precisely find which are the two entangled pure states in the $(2,2)$ decomposition we take yet another density matrix in this plane
\beq
\rho_{M'}=p\ketbra{{M'}_3}+(1-p)\ketbra{{M'}_3^\perp}\ .
\eeq 
Beyond the $(2,2)$ decomposition, closer to the Bloch sphere surface, none of the zero-states is visible and therefore the missing space up to the Bloch sphere surface is filled up with $(0,2)$ decompositions.
The convex roof here behaves strictly convex which makes it difficult to find the convex curve which is minimal. In this particular case we have a tiny fraction of the Bloch sphere remaining to be substituted by $(0,2)$ decompositions. The tangle of these states are found in the interval 
$[\min\{\tau[M(\varphi)],\tau[M'(\varphi)]\},\max_C\{\tau(M_C)\}]$, where $C$ is the intersection line of the Bloch sphere surface with the plane given by $M$, $M'$, and the two zero-states.
Future investigations may show how far this purely convex part of the convex roof can penetrate into the Bloch sphere.

For the remaining points in the Bloch sphere there is only a single visible state from the zero polytope left
which leads to $(1,1)$ decomposition being optimal there. They are depicted in Fig.~\ref{fig:OptimDecomp-GHZ-W} by green lines and red points.

\bibliography{refs.bib}

\end{document}